\catcode`\@=11
\def\gsim{\ifmmode{\mathrel{\mathpalette\@versim>}}
    \else{$\mathrel{\mathpalette\@versim>}$}\fi}
\def\lsim{\ifmmode{\mathrel{\mathpalette\@versim<}}
    \else{$\mathrel{\mathpalette\@versim<}$}\fi}
\def\@versim#1#2{\lower 2.9truept \vbox{\baselineskip 0pt \lineskip 
    0.5truept \ialign{$\m@th#1\hfil##\hfil$\crcr#2\crcr\sim\crcr}}}
\catcode`\@=12
\newcommand{\beq}{\begin{equation}}
\newcommand{\eeq}{\end{equation}}

\def\Lsun{L_{\odot}}
\def\Msun{M_{\odot}}

\def\eps{\epsilon}

\def\lb{L_{\rm B}}
\def\lbh{L_{\rm BH}}
\def\lx{L_{\rm X}}

\def\ledd{L_{\rm Edd}}

\def\lbhefUV{L_{\rm BH,UV}^{\rm eff}}
\def\lbhefopt{L_{\rm BH,opt}^{\rm eff}}

\def\lir{L_{\rm IR}}

\def\mast{M_*}

\def\mbh{M_{\rm BH}}

\def\re{R_{\rm e}}

\def\t15{t_{15}}

\def\Lsun{L_{\odot}}
\def\Msun{M_{\odot}}

\def\eps{\epsilon}

\def\epsw{\eps_{\rm w}}
\def\epswM{\epsw^{\rm M}}

\def\epsA{\eps_{\rm ADAF}}

\def\lb{L_{\rm B}}
\def\lbh{L_{\rm BH}}
\def\lx{L_{\rm X}}

\def\ledd{L_{\rm Edd}}

\def\lbhefUV{L_{\rm BH,UV}^{\rm eff}}
\def\lbhefopt{L_{\rm BH,opt}^{\rm eff}}

\def\lir{L_{\rm IR}}

\def\mast{M_*}

\def\mbh{M_{\rm BH}}

\def\re{R_{\rm e}}

\def\BIIzd{B2$_{02}$}

\def\BIIIzd{B3$_{02}$}
\def\BIIIzdl{B3$_{02}^{\rm l}$}
\def\BIIIzdh{B3$_{02}^{\rm h}$}

\def\t15{t_{15}}

\def\eps{\epsilon}

\def\epsw{\eps_{\rm w}}

\def\lbh{L_{\rm BH}}

\def\ledd{L_{\rm Edd}}

\def\Msun{M_{\odot}}

\documentclass[graybox]{svmult}

\usepackage{mathptmx}       
\usepackage{helvet}         
\usepackage{courier}        
\usepackage{type1cm}        
\usepackage{makeidx}         
\usepackage{graphicx}        
\usepackage{multicol}        
\usepackage[bottom]{footmisc}

\makeindex             

\begin{document}

\title*{AGN feedback in elliptical galaxies:\\ numerical simulations}
\author{Luca Ciotti and Jeremiah~P. Ostriker}
\institute{L. Ciotti \at Dept. of Astronomy, University of Bologna, via 
Ranzani 1, Bologna, 40127, Italy), \email{luca.ciotti@unibo.it}
\and J.P. Ostriker \at Princeton University Observatory, Peyton Hall, 
Princeton, NJ 08544-1001, USA, \email{ostriker@princeton.edu}\\
IoA, University of Cambridge, Madingley Road, Cambridge, CB3 0HA, UK}
%
%
\maketitle


\abstract{The importance of feedback (radiative and mechanical) from
  massive black holes at the centers of elliptical galaxies is not in
  doubt, given the well established relation among black hole mass and
  galaxy optical luminosity.  Here, with the aid of high-resolution
  hydrodynamical simulations, we discuss how this feedback affects the
  hot ISM of isolated elliptical galaxies of different mass.  The
  cooling and heating functions include photoionization plus Compton
  heating, the radiative transport equations are solved, and the
  mechanical feedback due to the nuclear wind is also described on a
  physical basis; star formation is considered.  In the medium-high
  mass galaxies the resulting evolution is highly unsteady.  At early
  times major accretion episodes caused by cooling flows in the
  recycled gas produced by stellar evolution trigger AGN flaring:
  relaxation instabilities occur so that duty cycles are small enough
  to account for the very small fraction of massive ellipticals
  observed to be in the QSO-phase, when the accretion luminosity
  approaches the Eddington luminosity. At low redshift all models are
  characterized by smooth, very sub-Eddington mass accretion
  rates. The mass accumulated by the central black hole is limited to
  range observed today, even though the mass lost by the evolving
  stellar population is roughly two order of magnitude larger than the
  black hole masses observed in elliptical galaxies.}

\section{Introduction}
\label{sec:1}

Supermassive black holes (SMBHs) at the centers of bulges and
elliptical galaxies (e.g., see \cite{KR95,DZ01,FF05}) have certainly
played an important role in the processes of galaxy formation and
evolution (e.g., see among others
\cite{SR98,Fabi99,BS01,CV02,Kin03,WL03,HCO04,
  Grana04,SOCS,MQT05,DMSH05,BN05,Hop06,Croto06,PSM09,LC10}), as
indicated by the observed correlations between host galaxy properties
and the masses of their SMBHs (e.g., see
\cite{Mag98,FM00,Geb00,YT02,McLD02, Grah03,MH03}, see also
\cite{Som08,Cat09,Ci09a}).

Of central interest for modern astrophysics is the fact that when gas
is added to the central galactic regions for any reason, the SMBH will
accrete and emit energy, both as a radiation flow and in some
mechanical form.  The complex interaction of such energy with the
galactic gas, and the consequent effects on the galaxy and on the SMBH
itself, are defined as ``AGN feedback''. A quite widespread view is
that, after the end of the galaxy formation epoch, the only way to add
fresh gas to the central SMBH is through the merging phenomenon; it
follows that the quasar phenomenon should be a secure indicator of
(gas rich) galaxy merging over the cosmic epochs.  However, this
picture is only partially true, as well known to stellar evolutionists
and to the ``cooling flow'' community.

In fact, the mass return rate from the passively evolving stars
(primarily from red giant winds and planetary nebulae) in elliptical
galaxies can be estimated as
\beq 
{\dot\mast}(t)\simeq
1.5\,10^{-11}\lb\,\t15^{-1.3} \quad \Msun {\rm
  yr}^{-1}, 
\eeq
where $\lb$ is the present galaxy blue luminosity in blue solar
luminosities, and $\t15$ is time in 15 Gyr units (\cite{CDPR}, see
also Sect.~2.2 and Pellegrini, this volume).  This metal-rich,
recycled gas is the main ingredient of the so-called cooling-flow
model (originally developed for clusters, \cite{CB77,FN77}), that
provided the preliminary framework to the interepretation of X-ray
halos observed in elliptical galaxies (e.g., see \cite{CFT,SW87}
and Fabbiano, this volume).

However soon it was realized that at least two major problems were
faced by the classical cooling-flow scenario.  The first is a {\it
  luminosity problem}, i.e, the X-ray luminosity $\lx$ of local
ellipticals is inconsistent with the standard cooling flow model. In
fact, low-redshift elliptical galaxies with optical luminosity
$\lb\gsim 3\times 10^{10}\Lsun$ show a significant range in the ratio
of gas-to-total mass at fixed $\lb$, with values ranging from
virtually zero up to few \%, and most of that is seen in X-rays with
temperatures close to the virial temperatures of the systems (e.g.,
see \cite{MB03}).  The second, and even more severe problem, is the
{\it mass disposal} problem. In fact, from eq.~(1) it follows that the
evolving stellar population will inject in the galaxy, over a
cosmological time, a gas mass summing up $\approx 10-20\%$ of the
total stellar mass $\mast$. In the classical cooling flow scenario,
this gas flows and disappears at the galaxy center, but observations
ruled out the existence of such large masses at the center of
elliptical galaxies. Young stellar populations observed in the body of
ellipticals also cannot account for the total mass released, and
alternative forms of cold mass disposal (such as distributed mass
drop-out) are not viable solutions (\cite{Bin01}). The mass disposal
problem has been exhacerbated after the discovery of central SMBHs:
infact, the total mass of the recycled gas is two orders of magnitude
larger than the mass $\mbh$ of the central SMBH. In other words, {\it
  even in absence of merging, the pure passive stellar evolution
  injects in the galaxy an amount of gas that, if flowing to the
  center, would produce a SMBH $\simeq 100$ times more massive than
  what is observed}, $\mbh\simeq 10^{-3}\mast$ (e.g., \cite{Mag98}).

A (partial) solution to the luminosity and mass problems was proposed
in a first series of papers (\cite{LM87,DRCP,DFJ90,CDPR,PC98}), by
considering the effect of SNIa heating of the galactic gas, and
exploring the time evolution of gas flows by using hydrodynamical
numerical simulations.  It was found that SNIa input sufficed for low
and medium-luminosity elliptical galaxies to produce fast galactic
winds, so that the scatter in the $\lx$-$\lb$ diagram could be nicely
reproduced.  However, it was also found that more massive galaxies
should be in the high-luminosity, permanent cooling-flow regime, so
that for massive systems, putative hosts of luminous cooling flows,
the mass problem was still unsolved; in addition, if this gas is
accreted on the central SMBH, then a bright QSO should be observed in
all X-ray luminous elliptical galaxies. These considerations lead
naturally to the study of gas accretion on SMBHs at the center of
elliptical galaxies, to explore the possibility that radiative and
mechanical feedback due to accretion is the solution of the mass
disposal problem in cooling flow, and it is the explanation of the
maintenance of ``small'' SMBH masses in presence of very large amounts
of recycled gas, and of the shut down of QSO activity in massive
ellitpicals (e.g., see \cite{TB93,CO97,FCE06}).

In the past years, we dedicated several papers to the AGN feedback in
elliptical galaxies (\cite{CO97,CO01,OC05,SOCS,CO07,COP09,PCO09,COP10,
  Jang10,Shin10a,Shin10b,OCCNP10,NOC10}).  The current most
satisfactory models are {\it combined models}, i.e. models in which
both radiative and mechanical feedback effects are at work.  In
general, all the computed solutions are characterized by relaxation
oscillations (e.g., see \cite{Ost76,COS78,MCB09}), and we note that
nowadays, several observations support the finding that accretion on
central SMBHs is in fact a highly unsteady phenomenon (e.g., see
\cite{Mart04,GSP08,HH09}); in addition, evidences of AGN feedback have
been clearly detected in the hot gas of nearby elliptical galaxies
(e.g., see \cite{Jon02,RSI04,OSVK05,FOR07,DS08a}, see also Statler,
this volume).

From the astrophysical point of view, the emerging picture of the
evolution of an isolated, medium-high mass elliptical galaxy consists
of four main (repeating) stages.

Stage 1) After the end of the galaxy formation epoch, the galaxy
should be in a more or less quiescent phase. Planetary nebulae and
other sources of secondary gas, processed through stellar evolution,
inject fresh gas in the galaxy at a rate proportional to the stellar
density and with an energy due to the stellar motions which gurantees
that, when the gas is thermalized, it will be approximately at the
local ``virial'' temperature. Supernovae (Type Ia) are also
distributed like the stars and will tend to drive a mild wind from the
outer parts of the galaxy, with the inner parts being quite luminous
in thermal X-rays. This is a ``normal'' giant elliptical galaxy. Low
mass ellipticals instead can be found permanently in a state of
global, low-luminosity galactic wind.

Stage 2) In massive ellipticals, the gas in the dense inner part of
the galaxy is radiating far more energy than can be replaced by SNIa
and stellar motions, and thus a ``cooling catastrophe'' occurs with a
collapsing cold shell forming at $\approx 1$ kpc from the center. As
this falls towards the center, a starburst occurs, and the galaxy seen
as an ULIRG.  A radio jet may be emitted, but the AGN flare up is at
first heavily obscured and the central source will only be seen in
hard X-rays.

Stage 3) Gradually, the gas is consumed, as it is transformed to new stars,
and some of it is driven out in a strong wind by the combined effects
of feedback from the starburst and the central SMBH, which is now
exposed as an optical and then UV ``quasar'', complete with Broad Line
Region (hereafter BLR) wind, optically thick disc of gas, and young
stars.

Stage 4) As gas is used up or blown away, a hot cavity is formed at
the center of the system and, since a shock has propagated through
that volume, it is essentially like a giant supernova remnant and one
expects there to be particle acceleration and non-thermal radiation
from the central region (\cite{Jang10}). Then, gradually this hot
bubble cools and collapses and one returns to the normal elliptical
phase at Stage 1.

The paper is organized as follows. In Section 2 we briefly discuss
some class of models that have been studied in the past years,
focusing on the radiative or mechanical feedback effects, but not
both. In Section 3 we describe in detail the input physics of the
combined feedback models.  In Section 4 we present for the first time
a comparison of the effects of combined feedback on three galaxy
models of different mass, related to the Reference Model in
\cite{COP10}. Finally, in Section 5 we discuss the main results
obtained.

\section{Previous works}
\label{sec:2}

Due to the importance of the subject, to its implications in different
areas of observational and theoretical astrophysics, and to the fact
the the specific nature of AGN feedback is still not completely
understood, it is not surprising that a very large body of work has
been done on the subject.  In general, past investigations focused
separately on {\it purely radiative} or {\it purely mechanical}
feedback.  Here we briefly describe the main properties and
limitations of these two classes of models.

\subsection{Radiative  feedback}
\label{sec:2.2}

In the published book.

\subsection{Mechanical feedback}
\label{sec:2.1}

In the published book.

\section{Physical modeling}
\label{sec:3}

In this Section we summarize the implementation of the input physics
in our 1D code, used to compute the evolution of {\it combined
  feedback} models. We now have a more advanced code version, that
will be used in future investigations, and we are also working on a 2D
code with a multidimensional implementation of the input physics.

\subsection{Structure and internal dynamics of the galaxy}

In the published book.

\subsection{Passive stellar evolution: SNIa rate and stellar mass losses}

In the published book.

\subsection{Star formation and SNII heating}

In the published book.

\subsection{The circumnuclear disk and the SMBH accretion luminosity}

In the published book.

\subsection{The mechanical feedback treatment}

In the published book.

\subsection{Radiative heating and cooling}

In the published book.

\subsection{Radiation pressure}

In the published book.

\subsection{Hydrodynamical equations}

In the published book.

\section{Results}
\label{sec:4}

We now illustrate the main properties of model \BIIIzd~(discussed in
\cite{COP10}) and two variants obtained by increasing (\BIIIzdh) and
decreasing (\BIIIzdl) its central velocity dispersion, while keeping
the remaining input physics identical. The galaxy models are
constructed as described in Sect~.3.1, and their structural parameters
are given in Table 1.  For sake of comparison, we recall that for
model \BIIIzd~ the initial stellar mass is $\mast\simeq 2.9\times
10^{11}\Msun$, the Fundamental Plane effective radius $\re\simeq6.9$
kpc, and the central aperture velocity dispersion $\sigma_{\rm a}=260$
km s$^{-1}$. The initial mass of the central SMBH is assumed to follow
the present day Magorrian relation ($\mbh\simeq 10^{-3}\mast$), as it
is believed that the bulk of the SMBH mass is assembled during the
process of galaxy formation (e.g., \cite{HCO04, SOCS,LC10}), a process
which is not addressed with the present simulations.  Note that these
models are not appropriate as initial conditions for cosmological
simulations, because their parameters are fixed to reproduce nearby
early-type galaxies (at $z=0$), and also because of the outflow
boundary conditions imposed at the galaxy outskirts ($\sim 250$ kpc).

\begin{table}
  \caption{The structural parameters of model \BIIzd~and its low and high 
    mass variants, and the relevant mass budgets (discussed in Sect.~4.2) 
    at the end of the simulations. 
    Velocity dispersions are in km/s, effective radii in kpc, 
    luminosities are in $10^{10}\Lsun$, stellar masses in $10^{11}\Msun$. 
    In the logarithms, masses are in Solar Masses.}
\label{tab:1}       
%
%
\begin{tabular}{p{1.1cm}p{1.1cm}p{1.1cm}p{1.1cm}p{1.1cm}p{1.3cm}p{1.2cm}p{1.2cm}p{1.2cm}}
\hline\noalign{\smallskip}
Model&$\sigma_0$&$\re$&$\lb$&$M_*$&$\log\Delta\mbh$&$\log\Delta
M_*$&$\log\Delta M_{\rm w}$&$\log M_{\rm ISM}$ \\
\noalign{\smallskip}\svhline\noalign{\smallskip}
\BIIIzdl       & 240  & 5.77 & 3.78 & 2.04 & 8.36 & 9.22  & 10.21&9.13\\
\BIIIzd        & 260  & 6.91 & 5.03 & 2.87 & 9.06 & 10.22 & 10.31&9.34\\
\BIIIzdh       & 280  & 8.20 & 6.59 & 3.95 & 9.41 & 10.58 & 10.40&9.75\\
\noalign{\smallskip}\hline\noalign{\smallskip}
\end{tabular}
\end{table}
The initial conditions for the ISM are represented by a very low
density gas at the local thermalization temperature.  The
establishment of such high-temperature gas phase at early cosmological
times is believed to be due to a ``phase-transition'' when, as a
consequence of star formation, the gas-to-stars mass ratio was of the
order of 10\% and the combined effect of shocks, SN explosions and AGN
feedback became effective in heating the gas and driving galactic
winds (e.g., see \cite{RCDP93,OC05,DMSH05,Sprin05,JNB08}).

Important quantities associated with the model evolution are the mass
(luminosity) accretion weighted EM and mechanical efficiencies
\beq 
<\epsA >\equiv {\int \epsA\dot\mbh\, dt \over\Delta\mbh };\quad 
<\epsw >\equiv {\int\epsw\dot\mbh\, dt \over\Delta\mbh}
\eeq 
where $\Delta\mbh$ is the mass accreted by the SMBH over the time
interval considered. In addition to the time-averaged quantities
introduced above, we also compute the {\it number of bursts} of each
model (each burst being counted when $\lbh$ becomes larger than
$\ledd/30$), the total time spent at $\lbh \geq \ledd/30$
(bolometric), the total time spent at $\lbhefUV\geq 0.2\ledd /30$ (UV,
after absorption), and at $\lbhefopt\geq 0.1\ledd /30$ (optical, after
absorption). The two numerical coefficients take into account the
fraction of the bolometric luminosity used as boundary condition to
solve the radiative transfer equation in each of the two bands (Sect.~3).

\subsection{Luminosities}
\label{sec:4.1}

\begin{figure}
\hskip 1truecm
\includegraphics[angle=0.,scale=0.6]{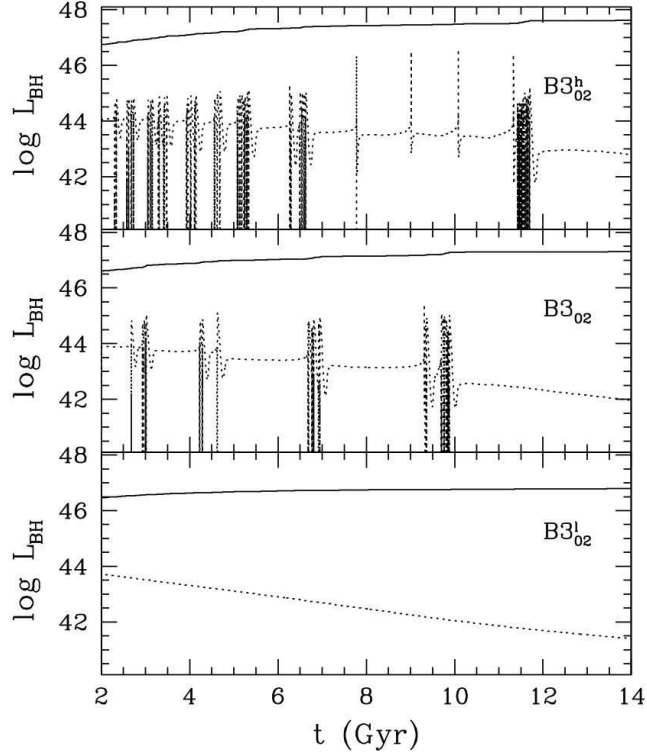}
\caption{Dotted lines are the optical SMBH luminosity corrected for
 absorption $\lbhefopt$ (i.e, as would be observed from infinity) for
 the three models.  We recall that at the center we fixed $\lbhefopt
 (R_1)=0.1\lbh$. The almost horizontal solid line is $\ledd$.  The
 structural properties of the galaxy models are given in Table 1. The
 feedback is of Type B, i.e. with a nuclear wind mechanical
 efficiency dependent on the (normalized) accretion luminosity
 $l\equiv\lbh/\ledd$, and with a peak mechanical efficiency of
 $\epswM=3\,10^{-4}$ and a peak radiative efficiency of $\eps_0=0.2$.
 The model in the central panel is discussed in detail in Paper
 III. (Adapted from \cite{COP10} by permission of the AAS).}
\label{lbh}
\end{figure}

The central panel of Fig.\ref{lbh} shows the evolution of the
accretion luminosity of model \BIIIzd, fully discussed in
\cite{COP10}.  After a first evolutionary phase in which a galactic
wind is sustained by the combined heating of SNIa and thermalization
of stellar velocity dispersion, the central ``cooling catastrophe''
commences. In absence of the central SMBH a ``mini-inflow'' would be
then established, with the flow stagnation radius (i.e., the radius at
which the flow velocity is zero) of the order of a few hundred pc to a
few kpc.  These ``decoupled'' flows are a specific feature of cuspy
galaxy models with moderate SNIa heating (\cite{PC98}).  However,
after the central cooling catastrophe, the feedback caused by
photoionization, Compton heating, and mechanical feedback, strongly
affects the subsequent evolution, as can be seen in Fig.~\ref{lbh}
where we show the luminosity evolution of the central AGN with
time-sampling of $10^5$ yrs. The corresponding Eddington limit is
represented by the almost horizontal solid line. As already discussed
in previous papers, the major AGN outbursts are separated by
increasing intervals of time (set by the cooling time and by mass
return rate from the evolving stellar population), and present a
characteristic temporal substructure, whose origin is due to the
cooperating effect of direct and reflected shock waves.  These
outflowing shocks are a likely place to produce emission of
synchrotron radiation and cosmic rays (\cite{Jang10,Sij08}). At
$t\simeq 10$ Gyr the SNIa heating, also sustained by a last strong
AGN burst, becomes dominant, a global galactic wind takes place and
the nuclear accretion switches to the optically thin regime.

The top and the low panels show instead the accretion luminosity for
the galaxy models with higher (top panel) and lower (bottom panel)
velocity dispersion. The differences are apparent, and are in line
with energetic expectations.  In fact, it is well known that big
elliptical galaxies are more bound (per unit mass) than low mass
systems (as dictated by the Fundamental Plane and Faber-Jackson
relations), while the specific heating provided by SNIa is independent
of the galaxy mass. For this reason, in model \BIIIzdh~not only the
bursting activity begins earlier than in model \BIIIzd, but also lasts
longer. The opposite case is represented by model \BIIIzdl, where the
SMBH accretion is found, over all the evolution, in the highly
sub-Eddington (ADAF), hot and optically thin regime, with absence of
central bursts. We note that the SMBH accretion luminosities of the
three models are far below the Eddigton limit at the current epoch, in
rough agreement with current observations, but clearly still more
luminous than the average low-luminosity objects (e.g., see
\cite{Pel05a,Ho08,Ho09,Pel10}). The need of an additional form of
feedback in the low-luminosity phases will be briefly addressed in the
Conclusions.

\begin{figure}
\hskip -0.6truecm
\includegraphics[angle=0.,scale=0.6]{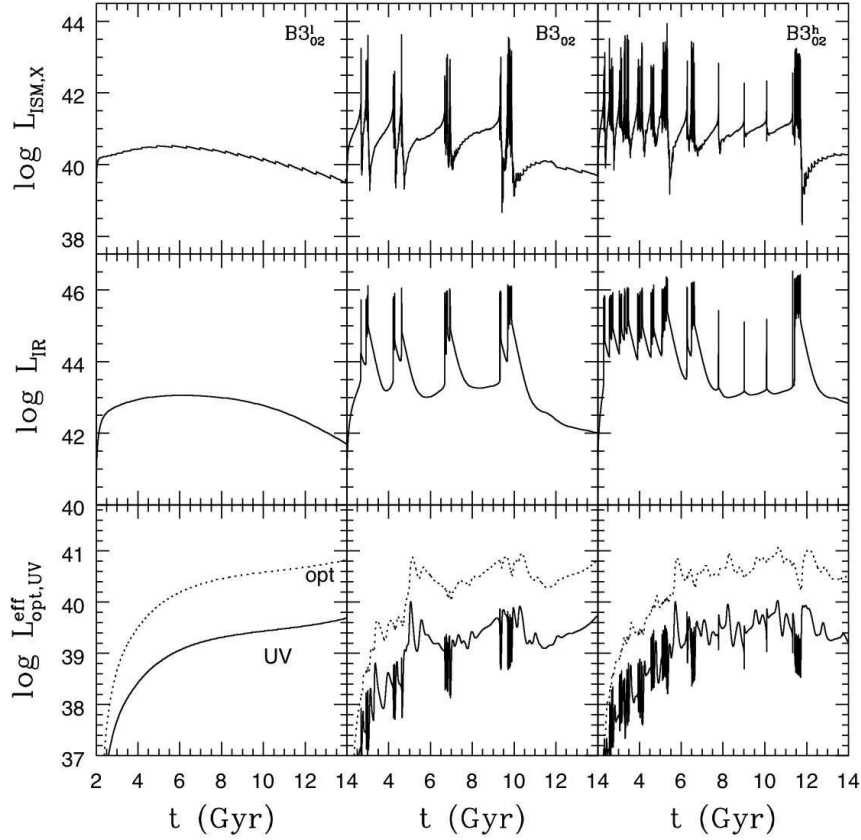}
\caption{Time evolution of the galaxy X-ray coronal luminosity $\lx$
  (top), recycled infrared $\lir$ (middle), and UV and
  optical luminosities (bottom), corrected for absorption.  The
  infrared luminosity is due to the reprocessing of the radiation
  emitted by the new stars and by the SMBH and absorbed by the ISM
  inside $10\re$. (Adapted from \cite{CO07} by permission of the AAS).}
\label{lum}
\end{figure}

In the top panels of Fig.~\ref{lum} we show the coronal X-ray
luminosity $\lx$ (emitted by gas at $T\geq 5\times 10^6$ K), due to
the hot galactic atmosphere integrated within $10\re$ for the three
models. Of course, in model \BIIIzdl~no bursts are observed,
consistently with the smooth nuclear accretion regime.  Instead, in
the other two models the spikes in the X-ray luminosity are clearly
reminescent of the SMBH accretion history.  These peaks are due to
sudden increases in the X-ray surface brightness profiles in the
central regions ($\approx 100$ pc scale), consequence of AGN feedback.
This is apparent from inspection of Figs.~6 and 7 (top left
panels). If the central regions are excluded from the computation of
$\lx$, this quantity would be seen to evelve in a much smoother way,
with fluctuations similar to those of the blu lines ($M_{\rm ISM}$) in
the top panels of Fig.~3. During more quiescent phases, $\lx$ attains
values comparable to the observed ones, with present times mean values
of $\lx$ lower than in the standard ``cooling flow'' model: it is
expected that a central galaxy in a cluster will reach higher values,
due to confining effects of the ICM, while stripping effects of the
ICM in satellite galaxies will lead to a further reduction
(\cite{Shin10b}).  Curiously, the $\lx$ values at the end of the
simulations are comparable. Of course, only a systematic exploration
of the parameter space determining the galaxy structure can confirm if
this is a robust result or just a fortuitous coincidence.  The most
natural explanation of the similarity of the $\lx$ values is that
$\lx$ of models \BIIIzd and \BIIIzdh has been finally reduced by the
series of bursts (absent in model \BIIIzdl): in fact, note how the
interburst $\lx$ of the two models is much higher than in the low-mass
model.  In the middle panels we show instead the estimated IR
luminosity $\lir$ due to the reprocessing of the radiation emitted by
the new stars and by the SMBH and absorbed by the ISM inside $10\re$.
Again, in the low mass model only a smooth evolution is
visible. Instead, in the other two models the bulk of the reprocessed
radiation comes from AGN obscuration, while the lower envelope is
determined by radiation reprocessing of the new stars.  Note that the
values of high luminosity peaks ($\lir\sim 10^{46}$ erg s$^{-1}$, or
more) are similar to those reported for ULIRGs (e.g., see
\cite{Pope06a,nard10}). In addition, peaks of nuclear IR emission
coupled with nuclear radio/X-ray emission have been recently reported
in a sample of elliptical galaxies (\cite{Tang}).  The bottom panels
present the temporal evolution of the optical and UV luminosities of
the new stars (corrected for absorption). A large fraction of the
starburst luminosity output (in the bursting models) occurs during
phases when shrouding by dust is significant (e.g., see
\cite{Rodi07,Brusa09}). At the end of the burst phase, the new stars
in the central regions will emit in UV and optical for $\approx 10^7$
yr, in the range seen in bright E+A sources. Nowadays, the different
time-scales of nuclear accretion and associated star formation can be
measured, with very interesting results (\cite{WHC10}).

As anticipated, we compute the duty-cycle as the total time spent by
the AGN at high luminosity phases, normalized to the age of the system
at the specified time.  In practice, we estimate the observable
duty-cycle as the fraction of the total time that the AGN is in the
``on'' state. The resulting values are very similar to the
luminosity-weighted values. First, the low-mass model, consistently
with the absence of bursts, has a null duty-cycle in the different
bands. Cumulative duty-cycles (i.e, spanning the whole simulation
time) of model \BIIIzd~are $\simeq (4.8\,10^{-2}, 2.7\,10^{-2},
1.9\,10^{-2})$, in the bolometric, optical and UV after absorption.
As expected (at each time) the larger duty-cycles are in the
bolometric, followed by absorbed optical and finally by absorbed
UV. Values for the more massive \BIIIzdh~are $\simeq (7.9\,10^{-2},
3.6\,10^{-2}, 2.3\,10^{-2})$. By construction these values cannot take
into account the temporal decline of the accretion activity over the
Hubble time.  For example, by restricting the computation to the
temporal baseline of the last 6 Gyr, the resulting duty-cycle values
drop by an order of magnitude.  These values compare nicely with
observational estimates (e.g., see \cite{Heck04,GrH07}).

\subsection{Mass budgets: SMBH, ISM, and starformation}
\label{sec:4.2}

\begin{figure}
\hskip -0.5truecm
\includegraphics[angle=0,scale=0.6]{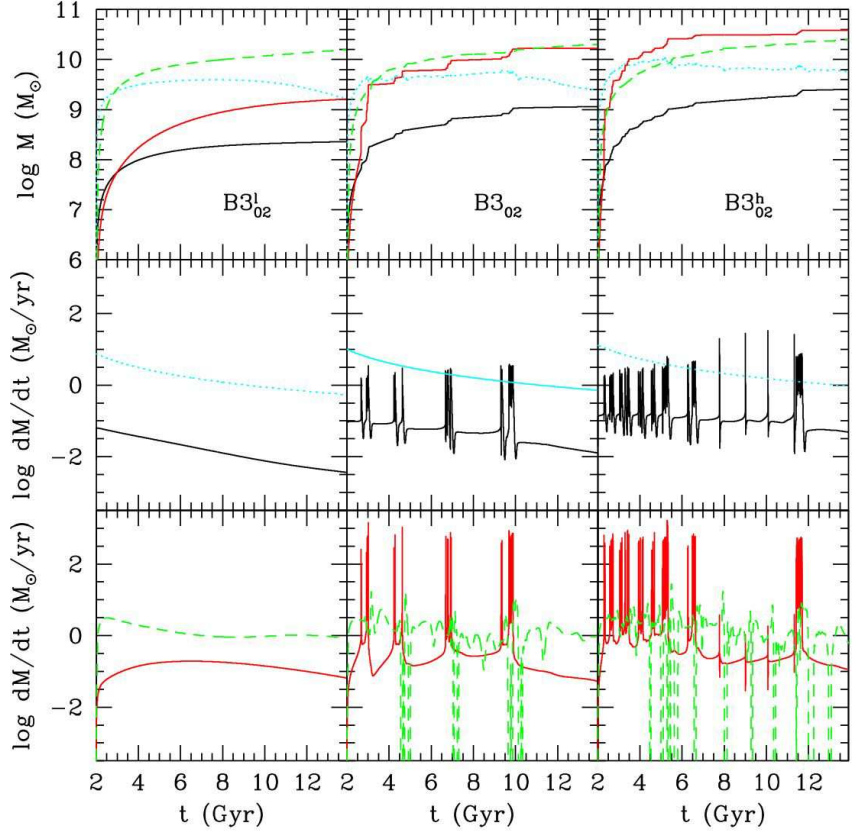}
\caption{Mass budget evolution of the models \BIIIzdl~(left column),
  \BIIIzd~(central column), and \BIIIzdh~(right column).  Top panels:
  total hot gas mass in the galaxy (within $10\re$, $M_{\rm ISM}$,
  blue lines), accreted mass on the central SMBH ($\Delta\mbh$, black
  lines), mass lost as a galactic wind at $10\re$ ($\Delta M_{\rm w}$,
  green lines), and total mass of new stars ($\Delta M_*$, red lines).
  Middle and bottom panels: time rates of the quantities in the top
  panels, with corresponding colours.}
\label{mass}
\end{figure}

In Fig.~\ref{mass} we show the time evolution of some of the relevant
mass budgets of the models (summarized in Table 1), both as
time-integrated properties and instantaneous rates: black lines refer
to the SMBH accretion ($\Delta\mbh$), green lines to the gas mass
ejected as a galactic wind ($\Delta M_{\rm w}$), red lines to the new
stars ($\Delta M_*$), and finally the blu lines to the gas content in
the galaxy. Of course, the SMBH accretion rate parallels the
luminosity evolution discussed in the previous Section.  A few
expected trends are apparent. For example, from the top panels it
results that the final accumulated SMBH mass is higher in the more
massive models. This is due to two reasons: first because the mass
return from the evolving stars in a galaxy scales linearly with the
stellar mass, and second, because the gas is more bound (per unit
mass) in more massive systems. The total mass ejected as a galactic
wind increases with the galaxy mass, but the remarkable fact here is
the strong dependence of the star formation history from the galaxy
mass. This is due, as already found and described in previous papers,
by the fact that in our models star formation is actually stimulated
by peak AGN activity. Therefore, AGN activity not only quenches star
formation (during the low-luminosity accretion phases), but it can
also be a trigger, especially during the ``passive'' evolution of
early-type galaxies. In any case, star formation episodes end abruptly
after major SMBH outburts.
The coincidence of vigorous star formation episodes with accretion
events and AGN activity can be clearly seen from the middle and bottom
panels, by comparison of the black and red lines. Note also how the
peaks in the green lines (galactic wind mass loss rate, $\Delta M_{\rm
  w}$)) are temporally displaced with respect to the starburst-AGN
episodes, due to the sound crossing time in the galaxy.  About the
galaxy mass loss, it is also important to note that the bulk of the
degassing is {\it not} due to AGN feedback events, but to the secular
heating provided by SNIa: absent this ingredient, all galaxy models
host gas inflows, with the consequent series of accretion events and
final SMBH masses well above the observed range.
\begin{figure}
\hskip 1.5truecm
\includegraphics[angle=0.,scale=0.4]{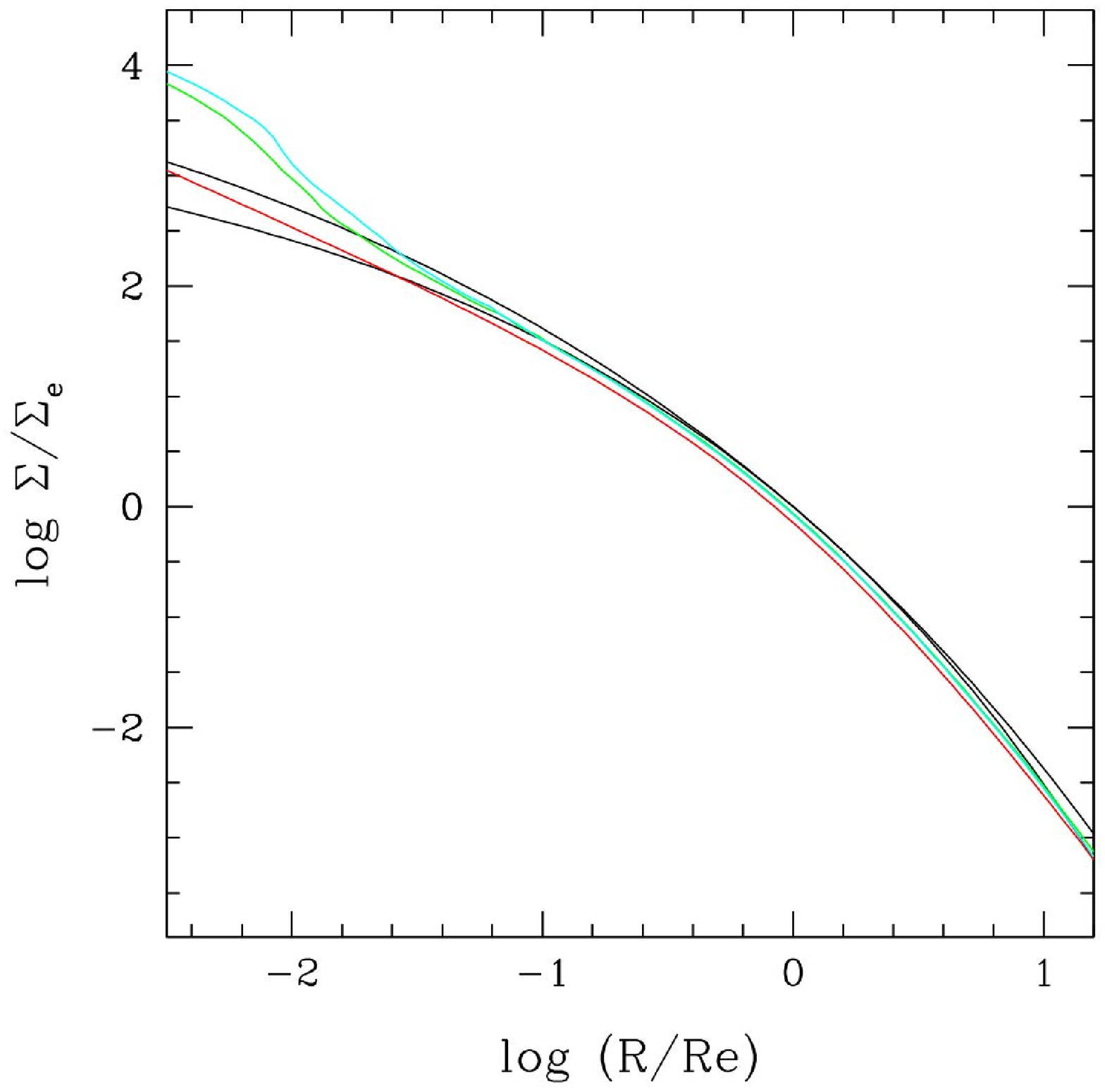}
\caption{Final stellar projected surface density profiles of model
  \BIIIzdl~(red), \BIIIzd~(green, see Paper III), and
  \BIIIzdh~(blue). Each profile is normalized to the surface density
  at the effective radius, while the radius is normalized to the
  effective radius.  The two black lines are the normalized global
  best-fit Sersic profiles of the initial and final projected
  profiles, with best-fit Sersic index $m\simeq 4.5$ and $m\simeq 6$,
  respectively. (Adapted from \cite{CO07,COP10} by permission of the
  AAS).}
\label{surf}
\end{figure}

As already mentioned above, these violent star formation episodes are
induced by accretion feedback\footnote{However, bursting star
  formation is not necessarily associated with AGN feedback
  (\cite{KT93}).}, and are spatially limited to the central 10-100
pc; thus, the bulk of gas flowing to the center is consumed in the
starburst. It is then expected that the final surface brightness
profile of the galaxy will be modified. In fact, this can be seen in
Fig.~\ref{surf}, where we show the final projected stellar density
profile of the models, together with Sersic (\cite{Ser68,CB99}) best-fit of
the initial and final profiles
\beq 
\Sigma(R)=\Sigma_0{\rm e}^{-b(R/\re)^{1/m}},\quad
b = 2m-1/3+4/405m+{\cal O}(m^{-2}).
\eeq 
The profiles show an increase with time of the best-fit Sersic
parameter $m$, from $\simeq 4.5$ up to $m\simeq 6$, within the range
of values commonly observed in ellipticals: also, in the final
\BIIIzd~and \BIIIzdh~models we note the presence of a central nucleus
originated by star formation which stays above the best fit profile.
Without entering the debated field of the morphological classification
of the centers of elliptical galaxies (e.g., see
\cite{Fab97,Grah04,GD05,Lau05, Davi07,Kor09,Shap10}, see also
\cite{Ci09b}), we notice that the ``light spikes'' in our models are
strikingly similar to the light spikes characterizing ``nucleated'' or
``extra-light'', and that usually are attributed to galaxy merging
(e.g., see \cite{Hop09}), and references therein). Observational
evidence is also accumulating that the central parts are quite metal
rich (e.g., see \cite{Chili09,Lee10} and, as noticed in \cite{Lau05},
where colors and luminosities of the nuclear regions of elliptical
galaxies are studied, on average the ``nuclear'' clusters are bluer
than the surrounding galaxy, as would be expected if the origin were
from infalling gas recycled from evolving stars. Finally, several
observational indications exist that, while the majority of the
stellar mass in elliptical galaxies may have formed at high redshifts,
small but detectable star formation events (summing up to $\lsim
5-10\%$ of the total stellar mass) have occurred at low redshift
(e.g., see \cite{WKI08,Pipi09,HWG08,TFD08}).

\subsection{Hydrodynamics}
\label{sec:4.3}

In the published book.

\section{Conclusions}
\label{sec:5}

In this review we have summarized the main results of combined
(radiative and mechanical, i.e., produced by direct interaction of a
nuclear wind/jet with the ISM) AGN feedback in elliptical galaxies,
obtained with the aid of high-resolution 1D hydrodynamical simulations
with a physically based feedback description. We presented for the
first time a comparison of feedback effects on galaxy models of
different mass. For completeness, we recall the main secure points on
which our framework is based.

First, it is known from stellar evolution theory, and supported by
observations, that the recycled gas from dying stars, available {\it
  independently of external phenomena such as galaxy merging}, sums up
to 20-30\% of the total mass in stars, and it is released over the
cosmic epoch. Therefore, recycled gas is an important source of fuel
for the central SMBH, with a total mass $\approx$ 2 orders of
magnitude larger than the mass measured in SMBHs in the local
universe.

Second, the metal rich recycled gas, if not removed from the parent
galaxy (by SNIa heating, ram-pressure, or tidal stripping), is
necessarily a subject of a classical radiative cooling instability,
leading to a collapse towards the center. This is the idea behind the
well known (and much debated) ``cooling flow'' scenario.

Third, as the cooling gas cannot disappear, a star-burst must occur
and also the central SMBH must be fed. The details of how much is
accreted on the central SMBH vs. consumed in stars vs. ejected from
the center by energy input from the starburst and the AGN are
uncertain.  But the observed mass of central SMBHs, and the mass of
the X-ray emitting hot gas, force to conclude that the bulk is
transformed into stars or blown out as a galactic wind, with less than
$1\%$ going into the central SMBH.

Fourth, since at the end of a major outburst a hot bubble remains at
the galaxy center, feedback processes shut themselves off, with a
recurrence time determined by stellar evolution and ISM cooling time.
Steady accretion on SMBHs is only possible at very low Eddington
ratios, and no steady flow appears to be possible for Eddington ratios
above $\simeq 0.01$.  Whenever the luminosity is significantly above
this limit, both the accretion and the output luminosity is in burst
mode.

Fifth, during the bursting phase the galaxy center would be optically
thick to dust, so one would observe a largely obscured starburst and a
largely obscured AGN, with most radiation in the far IR. As gas and
dust are consumed, the central source becomes visible. Much of the AGN
output occurs during obscured phases; then there is a brief interval
when one sees a ``normal'' quasar, and finally one would see a low
X-ray luminosity and E+A spectrum galaxy, in the central several
hundred pc, for $10^{7-8}$ yrs (e.g., \cite{GO07}).

All the simulations performed so far confirmed these expectations,
and the general results can be summarized as follows:

1) Radiative heating and radiation pressure on the ISM by photons
emitted by the central AGN and by the starburst, without any
mechanical input, greatly reduces the ``cooling flow catastrophe''
problem, but leads to a central SMBH that would be too bright and too
massive, and the galaxy would be too blue, due to repeated bursts of
central star formation.

2a) In absence of radiative feedback, mechanical energy from an AGN
wind with fixed efficiency also does not give a solution that in
detail satisfies the observations. For large efficiencies a giant
burst and an explosive degassing of the galaxy occurs (e.g.,
\cite{DMSH05, JNB08}). The gas content of the galaxy drops to levels
below what is observed in real elliptical galaxies and the systems
would have coronal X-ray luminosities orders of magnitude lower than
those typically seen in nearby ellipticals. Also, the computed AGN
duty cycle is too small. If the fixed efficiency is made low enough to
avoid these problems, then one reverts to the classical cooling flow
picture.

2b) Models with mechanical energy efficiency proportional to the
accretion luminosity, as indicated both by observations and detailed
2D hydrodynamical simulations for radiatively driven winds (e.g., see
\cite{KP08,KP09,KPN09}) perform better, but are still inadequate.  We
thus conclude that mechanical energy input - by itself - is unable to
provide appropriate levels of feedback that would leave ellipticals at
the current epoch with the properties that they are observed to have.

3) The combined models, in which both radiative and mechanical
feedback are allowed (as supported by observations, e.g.,
\cite{Alex10}), are the most satisfactorily.  This family of models,
with mechanical energy efficiency proportional to the luminosity, when
combined with a physically based treatment of the radiative effects,
does seem to be consistent with all observations for a range of
realistic efficiencies $\epsw$ (e.g., see \cite{Sij08}).  Radiative
and mechanical feedback affect different regions of the galaxy at
different evolutionary stages. During the ``quiescent'', optically
thin phases, radiative heating is distributed over all the galaxy
body, while the mechanical feedback is deposited in a region of a kpc
scale radius.  During the bursts, the collapsing cold shells are
optically thick, and most of the radiation is intercepted and
re-radiated in the IR; mechanical feedback plays a major role in
controlling accretion.

4) In combined models, radiative feedback from the central SMBH
(primarily the X-ray component) and the young star feedback consequent
to central star bursts (e.g., see \cite{TQM05}) can balance and
consume the cooling flow gas over the $10^2$-$10^3$ pc scale, but they
will not sufficiently limit the growth of the central SMBHs.
Mechanical feedback from the central SMBH on the $10$-$10^2$ pc scale,
mediated by the Broad Line Region winds (e.g., see
\cite{BT95,BN05,BR05,DMSH05}), is efficient in limiting the growth of
the SMBH, but, absent the radiative feedback, would leave elliptical
galaxies with more central star formation than observed.

From cosmological point of view, one of the main results of our study
is that {\it the evolution of an isolated galaxy, subject to
  internal evolution only, naturally leads to significant AGN and
  starburst activity, even in absence of external phenomena such as
  galaxy merging}. This conclusion is gaining more and more
observational support (e.g., see
\cite{Pier07,Li08,KH09,Tal09,Cister10}).

\subsection{Open questions and future developments}
\label{sec:5.1}

The investigation conducted so far, and summarized in the previous
Sections, suffers from a few weak points, namely: 1) the newly formed
stars are placed in the galaxy where they form; 2) the modifications
of the galaxy structure, gravitational field and velocity dispersion
profile, due to the stellar mass losses, galactic wind, and star
formation, are ignored; 3) the simulations are spherically symmetric,
so that Rayleigh-Taylor and Kelvin-Helmolthz unstable configurations
of the ISM (such as the formation of the cold shells, and the nuclear
wind and jet propagation), cannot be followed in detail.

The first two points will be addressed in future works. Instead, we
already started the exploration of 2D models, with very encouraging
and interesting results (\cite{NOC10}). Additional lines that have
been or will be studied are, for example, the properties expected for
the starburst population (such as spatial distribution, spectral
properties, etc.), the X-ray properties of the perturbed ISM as a
function of the combined effect of SNIa and central feedback
(\cite{PCO09}), and the cosmic rays emission following a central burst
(\cite{Jang10}).  Other obvious issues are the effects of environment,
as for a cD galaxy in a cluster, the stripping effects
(\cite{Shin10b}), and the impact of combined feedback models on the
ICM (extending the preliminary investigation \cite{COP04}, see also
\cite{Pope06b}).  We finally mention another observational riddle that
could be solved by the present models (with some additional work in
the physical modelization of feedback in the very sub-Eddington
accretion regime), i.e., that of the apparent ``underluminosity'' of
SMBHs in the local universe (e.g., see \cite{FC88,Pel05a}). In fact,
the simulations show clear evidence that an additional form of
feedback is needed during the quiescent, low-luminosity accretion
phases (in particular at late epochs). Of course, standard radiative
feedback is not effective during such phases, and presumably the
further reduction is provided by nuclear jets and/or thermally driven
winds (e.g., see \cite{Allen06,MH07}).
\begin{acknowledgement}

  We thank Ena Choi, Janfey Jang, Greg Novak, Silvia Pellegrini,
  Daniel Proga, Sergei Sazonov, Min-Su Shin, Anatoly Spitkovski,
  Rashid Sunyaev for their precious collaboration in the research
  effort described in this paper.  We also thank Dong-Woo Kim and
  Silvia Pellegrini for organizing the Joint Discussion at the IAU
  General Assembly in Rio, and for editing this volume. L.C. is
  supported by the MIUR Prin2008.

\end{acknowledgement}

\end{document}